\documentclass[12pt]{article}
\usepackage{fullpage,amsmath,amsthm}

\theoremstyle{plain}
\newtheorem{theorem}{Theorem}

\newtheorem{lemma}[theorem]{Lemma}

\theoremstyle{definition}

\theoremstyle{remark}

\newcommand{\bhm}{\textsc{bhm}}

\newcommand{\norm}[1]{\Vert #1 \Vert}

\newcommand{\zo}{\{0,1\}}

\newcommand{\vect}[1]{\mathbf{#1}}

\newcommand{\COMMENT}[1]{}

\newcommand{\vx}{\vect{x}}
\newcommand{\prob}{\mbox{Prob}}
\newcommand{\CR}{{\cal{R}}}
\newcommand{\CM}{{\cal{M}}}

\newcommand{\CT}{{\cal{T}}}

\newcommand{\CD}{{\cal{D}}}

\newcommand{\ket}[1]{|#1\rangle}
\newcommand{\bra}[1]{\langle#1|}

\newcommand{\inp}[2]{\langle{#1}|{#2}\rangle} % inproduct, < | >

\begin{document}

\title{The one-way communication complexity of the Boolean Hidden Matching Problem}
\author{Iordanis Kerenidis\thanks{Supported in part by ACI Securit\'e Informatique SI/03 511 and ANR AlgoQP grants of hte French Ministry and in part by the European Commission under the Intergrated Project Qubit Applications (QAP) funded by the IST directorate as Contract Number 015848.}\\
CNRS - LRI\\
Universit\'e Paris-Sud\\
jkeren@lri.fr
\and Ran Raz\thanks{Part of this work was done when the second author
visited Microsoft Research, Redmond.}\\
Faculty of Mathematics\\
Weizmann Institute\\
ran.raz@weizmann.ac.il}
\maketitle

\begin{abstract}
We give a tight lower bound of $\Omega(\sqrt{n})$
for the randomized one-way
communication complexity of the Boolean
Hidden Matching Problem \cite{BJK04}.
Since there is a
quantum one-way
communication complexity protocol of
$O(\log n)$ qubits for this problem,
we obtain an exponential separation of quantum and classical
one-way communication complexity for partial functions.
A similar result was independently obtained by Gavinsky, Kempe, de Wolf \cite{GKdW06}.

Our lower bound is obtained
by Fourier analysis, using the Fourier coefficients inequality
of Kahn Kalai and Linial~\cite{KKL88}.
\end{abstract}

\section{Introduction}

Communication complexity is a central model of computation,
first defined by Yao in 1979~\cite{Y79}. It has found applications in
many areas of theoretical computer science.
%including Boolean circuit complexity, time-space
%tradeoffs, data structures, automata, formulae size, etc.
Numerous
examples of such applications can be found in the textbook of
Kushilevitz and Nisan \cite{KN97}.

A communication complexity problem is defined by three sets $X,Y,Z$
and a relation $\CR \subseteq X\times Y\times Z$. There are two
unconditionally powerful parties, Alice and Bob, who are given inputs
$x\in X$ and $y\in Y$, respectively. Alice and Bob exchange messages
according to a shared {\em protocol} over a channel, until Bob has
sufficient information to announce an output $z \in Z$ s.t.\ $(x,y,z)
\in \CR$. The {\em communication cost} of a protocol is the sum of the
lengths of messages (in bits) Alice and Bob exchange on the worst-case
choice  of inputs $x$ and $y$.  The {\em communication complexity} of
the problem $\CR$ is the cost of the best protocol that computes $\CR$
correctly.

In the \emph{one-way} variant of the model \cite{PS84,A96,KNR99}, Alice is allowed to send a single message to Bob, after which  he announces the outcome. Last, in the \emph{Simultaneous Messages Passing (SMP)} model, Alice and Bob cannot communicate directly, but instead, each of them sends a single message
to a third party called the ``referee'', who computes the
outcome based on the two messages.

Another important distinction has to do with the type of problem that
Alice and Bob try to solve. In the most natural setting, the problem
is a total Boolean function, meaning that Alice and Bob receive
inputs $x \in \zo^n$ and $y \in \zo^n$ and the goal is to compute a Boolean function $f(x,y)$, which is defined for all possible $(x,y)$.
In other cases, the problem is a partial function (or promise problem),
meaning that Alice and Bob receive only inputs that satisfy some
special property and compute a Boolean function $f(x,y)$. For example, Alice and Bob might receive sets
$S$ and $T$ with the property that either they are disjoint or their
intersection is half their size and the question is to figure out
which of the two cases it is. Last, the communication problem could be
a relation, meaning that for each input of Alice and Bob there could
be more than one right answer. For example, on inputs $x\in\zo^n$ and
$y\in\zo^n$ Alice and Bob need to output an index $i\in[n]$ such that
$x_i=y_i$
(if such an $i$ exists).
Note that a total Boolean function is a special case  of a
promise problem, which is a special case of a relation.

We can define different measures of communication complexity for a
problem $\CR$ depending on the allowed protocols. In a
\emph{bounded-error randomized} protocol
with error $\delta$, we allow Alice and Bob to have access
to \emph{public} random coins.
For any inputs $x,y$,
the outcome $z$ should be correct with probability at least
$1-\delta$, where the probability is taken over the public random
coins. The cost of a randomized protocol is the number
of bits Alice and Bob exchange in the worst-case.
The randomized communication complexity of $\CR$ (w.r.t. $\delta$)
is the cost of the optimal randomized protocol for $\CR$.

In the setting of {\em quantum communication complexity} \cite{Y93}, Alice and
Bob have qubits, some of which
are initialized to their respective inputs. In a communication round,
a player can perform a unitary operation on his/her
part of the qubits and send some of them to the other player. At the
end of the protocol Bob performs a measurement and decides on an
outcome. The outcome of the protocol should be correct with
probability of at least $1 - \delta$ (for any inputs $x,y$).
The quantum communication complexity of $\CR$ is the number of qubits
exchanged in the optimal bounded-error quantum protocol for $\CR$.

The main question in the theory of quantum communication complexity
is whether in the different communication models quantum channels can
reduce significantly the amount of communication necessary to solve
certain types of problems.

For total functions, we do not have any exponential gap between
quantum communication and randomized communication with public coins
in any of the abovementioned models.
Buhrman \emph{et al.\
}\cite{BCWW01} were able to solve the equality
problem in the SMP model with a quantum protocol of complexity $O(\log n)$
rather than the $\Theta(\sqrt{n})$ bits necessary in any bounded-error
randomized SMP protocol with private coins \cite{NS96,BK97}.
However, if we allow the players to share random coins, then equality
can be solved classically with $O(1)$ communication.

For promise problems, an exponential gap
between the quantum and the (public-coins)
randomized communication complexity models was proved in~\cite{R99}.
This was obtained by
describing a promise problem ${\cal{P}}_1$ with an efficient
quantum protocol of complexity $O(\log n)$ and such that
the bounded-error randomized communication
complexity of ${\cal{P}}_1$
is $\Omega(n^{1/4})$.

For relations, Bar-Yossef {\em et. al.} \cite{BJK04} defined the
Hidden Matching Problem and proved an exponential gap between quantum
and randomized communication in the one-way model and the SMP model.
%It was an open question to prove an exponential gap for a promise
%problem in the case of one-way communication.

In this paper we give a tight lower bound of $\Omega(\sqrt{n})$
for the bounded-error randomized one-way
communication complexity of a Boolean version of the
Hidden Matching Problem \cite{BJK04}.
Since there is a simple
quantum one-way
communication complexity protocol of
$O(\log n)$ qubits for the problem,
this provides an exponential separation for promise
problems between the models of quantum and randomized one-way
communication complexity.
A similar result was independently obtained by Gavinsky, Kempe, de Wolf \cite{GKdW06}.

Our lower bound is obtained
by Fourier analysis, using the Fourier coefficients inequality
of Kahn Kalai and Linial~\cite{KKL88},
which in turn was proved using the Bonami-Beckner
inequality~\cite{Bon70,Bec75}.
The KKL inequality was previously used
in the context of communication complexity in~\cite{Raz95,Kla01}.

\section{Preliminaries}

\subsection{Fourier analysis}

For a function $f:\zo^n \rightarrow \CR$, we define the $\ell_1$ and $\ell_2$ norms as
\[ ||f||_1 = \sum_{x\in\zo^n}|f(x)| \;\;\;, \;\;\; ||f||_2 = \left( \sum_{x\in\zo^n} |f(x)|^2 \right)^{1/2}
\]
It is a well known fact that for a function $f:\zo^n \rightarrow \CR$
\[ ||f||_2^2 \geq \frac{||f||_1^2}{2^n}.
\]
The Fourier transform of $f:\zo^n \rightarrow \CR$ is defined as
\[ f = \sum_{s\in\zo^n} \hat{f}(s)\chi_s,
\]
where $\chi_s:\zo^n\rightarrow \CR$ is the character $\chi_s(y)=(-1)^{y^T\cdot\;  s}$ with $``\cdot"$ being the scalar product over $GF(2)$ and $\hat{f}(s)$ is the Fourier coefficient
\[ \hat{f}(s) = \frac{1}{2^n}\sum_{y\in\zo^n} f(y)\chi_s(y).
\]
One very useful fact about the Fourier coefficients of a function  is Parseval's identity
\begin{lemma}(Parseval's Identity)\\
For a function $f:\zo^n \rightarrow \CR$ it holds that
\[ ||f||_2^2 = 2^n \sum_{s\in\zo^n} (\hat{f}(s))^2 \]
\end{lemma}
Let $f:\zo^n \rightarrow \CR$ and $g:\zo^n \rightarrow \CR$ and $``+"$ denote the bitwise XOR of two strings. The convolution $f\ast g:\zo^n \rightarrow \CR$ is defined as
\[ f\ast g(w) = \sum_{y\in\zo^n}f(y+w)g(y)
\]
For the Fourier coefficients of a convolution we have the following lemma
\begin{lemma}
For functions $f:\zo^n \rightarrow \CR$ and $g:\zo^n \rightarrow \CR$ it holds that
\[
\widehat{f\ast g}(s) = 2^n \cdot \hat{f}(s)\cdot \hat{g}(s)
\]
\end{lemma}
Let $h(\cdot,\cdot)$ be the hamming distance function and $h(\cdot)$ the hamming weight function. A final tool in our analysis is the KKL lemma
\begin{lemma}\cite{KKL88}
Let $f$ be a function $f:\zo^n\rightarrow\{-1,0,1\}$. Let $t$ be the probability that $f\neq 0$. Then for every $0\leq \delta \leq 1$
\[ \sum_{s\in\zo^n} \delta^{h(s)}(\hat{f}(s))^2 \leq t^{\frac{2}{1+\delta}}
\]
\end{lemma}

\subsection{Quantum computation}

We explain the standard notation of quantum computing and
describe the basic notions that will be useful in this paper. For more
details we refer the reader to the textbook of Nielsen and Chuang
\cite{NC00}.

Let $H$ denote a 2-dimensional complex vector space,
equipped with the standard inner product. We pick an orthonormal
basis for this space, label the two basis vectors
$\ket{0}$ and $\ket{1}$, and for simplicity identify them
with the vectors $\left(\begin{array}{c}1\\ 0\end{array}\right)$
and $\left(\begin{array}{c}0\\ 1\end{array}\right)$, respectively.
A \emph{qubit} is a unit length vector in this space, and so can be
expressed as a linear combination of the basis states:
$$
\alpha_0\ket{0}+\alpha_1\ket{1}=\left(\begin{array}{c}\alpha_0\\ \alpha_1 \end{array}\right).
$$
Here $\alpha_0,\alpha_1$ are complex \emph{amplitudes},
and $|\alpha_0|^2+|\alpha_1|^2=1$.

An \emph{$m$-qubit system} is a unit vector in the $m$-fold tensor space
$H\otimes\cdots\otimes H$.  The $2^m$ basis states of this space are
the $m$-fold tensor products of the states $\ket{0}$ and $\ket{1}$.
For example, the basis states of a 2-qubit system are the four 4-dimensional
unit vectors $\ket{0}\otimes\ket{0}$, $\ket{0}\otimes\ket{1}$,
$\ket{1}\otimes\ket{0}$, and $\ket{1}\otimes\ket{1}$.
We abbreviate, e.g., $\ket{1}\otimes\ket{0}$ to $\ket{1}\ket{0}$, or
$\ket{1,0}$, or $\ket{10}$, or even $\ket{2}$ (since 2 is 10 in binary).
With these basis states, an $m$-qubit state $\ket{\phi}$
is a $2^m$-dimensional complex unit vector
$$
\ket{\phi}=\sum_{i\in\zo^m}\alpha_i\ket{i}.
$$
We use $\bra{\phi}=\ket{\phi}^*$ to denote the conjugate transpose
of the vector $\ket{\phi}$, and $\inp{\phi}{\psi}=\bra{\phi}\cdot\ket{\psi}$
for the inner product between states $\ket{\phi}$ and $\ket{\psi}$.
These two states are \emph{orthogonal} if $\inp{\phi}{\psi}=0$.
The \emph{norm} of $\ket{\phi}$ is $\norm{\phi}=\sqrt{\inp{\phi}{\phi}}$.

Let $\ket{\phi}$ be an $m$-qubit state and
$B=\{\ket{b_1},\ldots,\ket{b_{2^m}}\}$ an orthonormal basis
of the $m$-qubit space. A measurement of the state $\ket{\phi}$ in the
$B$ basis means that we apply the projection operators
$P_i=\ket{b_i}\bra{b_i}$ to $\ket{\phi}$. The resulting quantum state
is $\ket{b_i}$ with probability $p_i=|\inp{\phi}{b_i}|^2$.

%%%%%%%%%%%%%%%%%%%%%%%%%%%%%%%%%%%%%%%%%%%%%%%%%%%%%%%%%%%%%
%%%%%%%%%%%%%%%%%%%%%%%%%%%%%%%%%%%%%%%%%%%%%%%%%%%%%%%%%%%%%
%%%%%%%%%%%%%%%%%%%%%%%%%%%%%%%%%%%%%%%%%%%%%%%%%%%%%%%%%%%%%
%%%%%%%%%%%%%%%%%%%%%%%%%%%%%%%%%%%%%%%%%%%%%%%%%%%%%%%%%%%%%

\section{Definition of the Boolean Hidden Matching Problem}

The relational version of the Hidden Matching Problem was defined in \cite{BJK04}. There, they proved a $\Omega(\sqrt{n})$ lower bound for the randomized one-way communication complexity of it and also described a $O(\log n)$ quantum one-way protocol. This provided an exponential separation for a relation between the randomized and quantum one-way communication complexity models. \cite{BJK04} also defined a Boolean version of the Hidden Matching Problem but did not provide a lower bound for the randomized communication complexity of it.

A version of the relational Hidden Matching Problem was also used by Gavinsky {\em et al} \cite{GKRdW06} to show that in the model of Simultaneous Messages, shared entanglement can reduce the communication exponentially compared to shared randomness.

Here, we define a slightly different version of the Boolean Hidden Matching Problem and prove a tight lower bound for its randomized one-way communication complexity.
We denote a perfect matching on $[2n]$ as a $(n \times 2n)$ binary matrix $M$ where each column corresponds to a number in $[2n]$ and the $i$-th row corresponds to the $i$-th edge of the matching.
In other words, if the $i$-th edge of the matching is $(k,l)$, then the $i$-th row of the matrix contains two 1's at the positions $k$ and $l$ and 0's elsewhere.

Let $x\in \zo^{2n}$. Then the product $Mx$ is an $n$-bit string $w$, where the $i$-th bit is equal to the parity of the two bits of $x$ that correspond to the $i$-th edge of the matching, i.e. $w_i = x_k \oplus x_l.$
Recall that we denote by $h(\cdot,\cdot)$ the hamming distance function and by $h(\cdot)$ the hamming weight function.  \\

\noindent
{\bf The Boolean Hidden Matching Problem (BHM$_n$):}

Alice gets as input a string $x \in \zo^{2n}$ and Bob gets as input a perfect matching $M$ on $[2n]$ and a string $w\in \zo^n$.
The promise is that either
$h(Mx,w) \leq n/3$ (``0'' instance) or $h(Mx,w) \geq 2n/3$ (``1'' instance). The goal is for Bob to determine where the input corresponds to a ``0'' instance or to a ``1'' instance.\\

\section{Quantum protocol for the Boolean Hidden Matching Problem}

We present a quantum protocol for the Boolean Hidden Matching Problem
with communication complexity of $O( \log n)$
qubits. Let $\vx = x_1\ldots x_{2n}$ be Alice's input and $M,w$ be
Bob's input. \\

\noindent
{\sc Quantum protocol for $\bhm_n$}
\begin{enumerate}
\item Alice sends the state $\ket{\psi}=\frac{1}{\sqrt{2n}}\sum_{i=1}^{2n}
(-1)^{x_i}\ket{i}$.
\item Bob performs a measurement on the state $\ket{\psi}$
in the orthonormal basis\\
$B=\{\frac{1}{\sqrt{2}}(\ket{k}\pm \ket{\ell}) \mid (k,\ell)\in M\}$. \\
\end{enumerate}

The probability that the outcome of the measurement is a basis
state $\frac{1}{\sqrt{2}}(\ket{k}+ \ket{\ell})$ is
\[ |\inp{\psi}{\frac{1}{\sqrt{2}}(\ket{k}+ \ket{\ell})}|^2=
\frac{1}{4n}((-1)^{x_k}+(-1)^{x_\ell})^2.\]
This equals to $1/n$ if
$x_k\oplus x_{\ell}=0$ and 0 otherwise.  Similarly for
the state $\frac{1}{\sqrt{2}}(\ket{k}- \ket{\ell})$ we have that
$|\inp{\psi}{\frac{1}{\sqrt{2}}(\ket{k}- \ket{\ell})}|^2$ is 0 if
$x_k\oplus x_{\ell}=0$ and $1/n$ if $x_k\oplus x_{\ell}=1$.
Hence, if the outcome of the measurement is a state
$\frac{1}{\sqrt{2}}(\ket{k}+
\ket{\ell})$ then Bob knows with certainty that $x_k\oplus
x_{\ell}=0$. If the outcome is a state
$\frac{1}{\sqrt{2}}(\ket{k}-\ket{\ell})$ then Bob knows with
certainty that $x_k\oplus x_{\ell}=1$. Let $(k,\ell)$ be the $j$-th
edge in the matching $M$, then Bob outputs $x_k\oplus x_{\ell} \oplus
w_j$. The protocol is correct with probability at least $2/3$ and by repeating
a constant number of times we can achieve correctness $1-\epsilon$ for
any small constant $\epsilon$.

%%%%%%%%%%%%%%%%%%%%%%%%%%%%%%%%%%%%%%%%%%%%%%%%%%%%%%%%%%%%%
%%%%%%%%%%%%%%%%%%%%%%%%%%%%%%%%%%%%%%%%%%%%%%%%%%%%%%%%%%%%%
%%%%%%%%%%%%%%%%%%%%%%%%%%%%%%%%%%%%%%%%%%%%%%%%%%%%%%%%%%%%%
%%%%%%%%%%%%%%%%%%%%%%%%%%%%%%%%%%%%%%%%%%%%%%%%%%%%%%%%%%%%%
%%%%%%%%%%%%%%%%%%%%%%%%%%%%%%%%%%%%%%%%%%%%%%%%%%%%%%%%%%%%%

\section{The randomized one-way communication complexity of the Boolean Hidden
Matching Problem}

\begin{theorem}
The randomized one-way communication complexity of the Boolean Hidden Matching
Problem is $\Omega(\sqrt{n})$.
\end{theorem}

\begin{proof}

For $b\in\zo$ we denote by $\sigma_b$ the distribution over $\zo$ such that $\prob_{\sigma_b}[b]=3/4$ and define $\mu_b = (\sigma_b)^{\otimes n}$. In other words, $\mu_0$ is the distribution over strings $\zo^n$ such that independently for every bit $i\in[n]$, $\prob_{\mu_0}[i \mbox{-th bit is 0}]= 3/4$ and $\mu_1$ is the distribution over strings $\zo^n$ such that independently for every bit $i\in[n]$, $\prob_{\mu_1}[i \mbox{-th bit is 1}]= 3/4.$

We define the function $f:\zo^{n}\rightarrow \cal{R}$ as
\[ f(y) =  \prob_{\mu_0}[y] - \prob_{\mu_1}[y] \]
It is easy to see that the Fourier coefficients of $f$ are
\[ \hat{f}(s) = \left\{ \begin{array}{ll} \frac{2}{2^{n+k}} &
\mbox{ for }s \mbox{ with } h(s)=k \mbox{ and } k \mbox{ odd}  \\ 0 &
\mbox{ otherwise}
\end{array} \right.\]

Using Yao's Lemma \cite{Y83}, in order to prove the lower bound, it suffices to
construct a ``hard'' distribution over instances of $\bhm_n$, and prove a lower bound for deterministic one-way protocols whose distributional error with respect to this distribution is at most $\epsilon$.\\
For every $x\in\zo^{2n}$ and matching $M$ we define the following distributions:\\
$\CD_0$ is the distribution over strings $w\in \zo^n$ such that independently for each $i\in [n]$, $\prob[w_i = (Mx)_i] = 3/4$ and $\CD_1$ is the distribution over strings $w\in \zo^n$ such that independently for each $i\in [n]$, $\prob[w_i \neq (Mx)_i] = 3/4$.

The "hard" distribution $\CT$ is defined as follows:
The string $x\in\zo^{2n}$ and the matching $M$ are picked uniformly at random. The string $w\in\zo^n$ is picked according to the distribution $\CD = \frac{1}{2}\CD_0 + \frac{1}{2}\CD_1$, that is $w$ is picked with probability $1/2$ from the distribution $\CD_0$ and with probability $1/2$ from the distribution $\CD_1$, where $\CD_0,\CD_1$ are the ones corresponding to $x,M$. The goal is now for Bob to determine whether $w$ was drawn from the distribution $\CD_0$ or $\CD_1$.

Note that if $(x,M,w)$ are picked according to the distribution $\CT$, then the probability that $n/3 \leq h(Mx,w)\leq 2n/3$ is exponentially small. Hence, any probabilistic protocol for $\bhm_n$ with error $\epsilon'$ gives a deterministic protocol for the distribution $\CT$ with distributional error $\epsilon'+o(1)$. Therefore, for the rest of the proof we use the distribution $\CT$.

Let us assume that there exists a deterministic protocol $P$ which is correct on the distribution $\CT$ with probability $1-\epsilon$, namely for uniformly random $x,M$ and $w$ drawn from $\CD$, Bob can determine with probability $1-\epsilon$ whether $w$ was drawn from $\CD_0$ or $\CD_1$.
Then, the protocol is correct for at least half of the $x$'s, with probability at least $(1-2\epsilon)$ over the input of Bob. Let us denote by $S \subseteq \zo^{2n}$ the set of these ``good" $x$'s.

Let $A \subset S$ be a set of $x$'s, for which Alice sends the same
message. Note that since $A$ is a subset of $S$, the protocol is correct with probability at least $(1-2\epsilon)$ over the inputs $x\in A$. We are going to show that the size of $A$ cannot be too large.

Let $g:\zo^{2n} \rightarrow \CR$ be the uniform
distribution over the set $A$, i.e.
\[ g(x) =  \left\{ \begin{array}{ll} \frac{1}{|A|} &
\mbox{ for } x\in A  \\
0 & \mbox{ for } x\not\in A \end{array} \right.  \]
For any matching $M$ we define $g_M:\zo^n \rightarrow \CR$ to be the distribution of $Mx$ when $x$
is picked uniformly from the set $A$, i.e.
\[ g_M(y) =  \frac{|\{ x\in A | Mx=y\}|}{|A|}  \]

For the $\ell_1$ norm of $f \ast g_M$ we have
\begin{eqnarray*}
||f \ast g_M ||_1 & = & \sum_{ w \in \zo^n} |f \ast g_M(w) |
\; = \; \sum_{ w \in \zo^n} \left| \sum_{ y \in \zo^n} f(y+w)g_M(y)\right| \\
& = & \sum_{ w \in \zo^n} \left| \frac{1}{|A|} \sum_{ x \in A} f(Mx+w) \right| \\
\end{eqnarray*}
where the last equation follows from the definition of the function $g_M$.
Since Alice's message is fixed for the set $A$ and Bob's algorithm is deterministic, for every matching $M$ we can split the set of $w$'s into two sets $W_{b,M}$, where $b\in\zo$ is Bob's answer. Let $\CM$ denote the set of all possible perfect matchings on $[2n]$ and $N = |\CM|$. Then
\begin{eqnarray*}
\lefteqn{\frac{1}{N} \sum_{M\in\CM} ||f \ast g_M ||_1}\\
& = & \frac{1}{N} \sum_{M\in\CM} \sum_{ w \in W_{0,M}} \left| \frac{1}{|A|} \sum_{ x \in A} f(Mx+w) \right|+
\frac{1}{N} \sum_{M\in\CM} \sum_{ w \in W_{1,M}} \left| \frac{1}{|A|} \sum_{ x \in A} f(Mx+w) \right|\\
& \geq & \frac{1}{N |A|} \left| \sum_{M\in\CM} \sum_{ w \in W_{0,M}} \sum_{x\in A} f(Mx+w)\right| +
\frac{1}{N |A|} \left| \sum_{M\in\CM} \sum_{ w \in W_{1,M}} \sum_{x\in A} f(Mx+w)\right| \\
& = & \frac{1}{N |A|} \left| \sum_{M\in\CM} \sum_{ w \in W_{0,M}} \sum_{x\in A} \left( \prob_{\mu_0}[Mx+w]- \prob_{\mu_1}[Mx+w]\right) \right| \\
& + &
\frac{1}{N |A|} \left| \sum_{M\in\CM} \sum_{ w \in W_{1,M}} \sum_{x\in A} \left( \prob_{\mu_0}[Mx+w] -\prob_{\mu_1}[Mx+w]\right)\right| \\
& \geq & \frac{1}{N |A|} \left| \sum_{M\in\CM} \sum_{ w \in W_{0,M}} \sum_{x\in A} \prob_{\mu_0}[Mx+w]\right| - \frac{1}{N |A|} \left| \sum_{M\in\CM} \sum_{ w \in W_{0,M}} \sum_{x\in A} \prob_{\mu_1}[Mx+w] \right| \\
& + & \frac{1}{N |A|} \left| \sum_{M\in\CM} \sum_{ w \in W_{1,M}} \sum_{x\in A} \prob_{\mu_1}[Mx+w] \right| - \frac{1}{N |A|} \left| \sum_{M\in\CM} \sum_{ w \in W_{1,M}} \sum_{x\in A} \prob_{\mu_0}[Mx+w] \right|
\end{eqnarray*}
The protocol is correct when Bob answers $b\in\zo$ and the string $w$ was drawn from the distribution $\CD_b$. Since the protocol is correct with probability at least $1-2\epsilon$ we have
\begin{eqnarray*}
\frac{1}{N}  \sum_{M\in\CM} \frac{1}{|A|} \sum_{x\in A} \sum_{ w \in W_{0,M}} \frac{1}{2}\prob_{\CD_0}[w]
+ \frac{1}{N}  \sum_{M\in\CM} \frac{1}{|A|} \sum_{x\in A} \sum_{ w \in W_{1,M}} \frac{1}{2}\prob_{\CD_1}[w] & \geq & 1- 2\epsilon\\
\frac{1}{N |A|}  \sum_{M\in\CM} \sum_{ w \in W_{0,M}} \sum_{x\in A} \prob_{\mu_0}[Mx+w]
+ \frac{1}{N |A|}  \sum_{M\in\CM} \sum_{ w \in W_{1,M}} \sum_{x\in A} \prob_{\mu_1}[Mx+w] & \geq & 2- 4\epsilon\\
\end{eqnarray*}
Similarly,
\begin{eqnarray*}
\frac{1}{N |A|}  \sum_{M\in\CM} \sum_{ w \in W_{0,M}} \sum_{x\in A} \prob_{\mu_1}[Mx+w]
+ \frac{1}{N |A|}  \sum_{M\in\CM} \sum_{ w \in W_{1,M}} \sum_{x\in A} \prob_{\mu_0}[Mx+w] & \leq & 4\epsilon
\end{eqnarray*}
Hence, we conclude that
\begin{eqnarray*}
\frac{1}{N} \sum_{M\in\CM} || f \ast g_M ||_1 & \geq & 2(1-4\epsilon)
\end{eqnarray*}
The $\ell_1$ and $\ell_2$ norms are related by the following inequality
\[ ||f\ast g_M ||_2^2 \geq \frac{||f\ast g_M||_1^2}{2^{n}} \]
and hence for the $\ell_2$ norm we have
\begin{eqnarray}
\frac{1}{N} \sum_{M\in\CM} ||f\ast g_M||^2_2 & \geq & \frac{1}{N} \sum_{M\in\CM} \frac{||f\ast g_M||^2_1}{2^{n}} = \frac{1}{2^{n}} \frac{1}{N}\sum_{M\in\CM} ||f\ast g_M||_1^2  \geq  \frac{1}{2^{n}} \left( \frac{1}{N}\sum_{M\in\CM} ||f\ast g_M||_1 \right)^2 \nonumber \\
& \geq & \frac{1}{2^{n-2}} (1-4\epsilon)^2 \label{1}
\end{eqnarray}
By Parseval's identity (lemma 1) and the convolution theorem (lemma 2) it holds that
\[ ||f\ast g_M||_2^2 = 2^n \sum_s (\widehat{f\ast g_M}(s))^2 = 2^{3n} \sum_s
( \hat{f}(s))^2 (\hat{g_M}(s))^2 \]
Using the expression for the Fourier coefficients of $f$ we have
\begin{eqnarray}\label{intermed}
\frac{1}{N}\sum_{M\in \CM}  ||f\ast g_M||^2_2 & = &
\frac{1}{N}\sum_{M\in \CM} 2^{3n} \sum_s (\hat{f}(s))^2(\hat{g_M}(s))^2 =
 \frac{2^{3n}}{N}\sum_{M\in \CM}  \mathop{\sum_{s:h(s)=k}}_{k\; odd}  \frac{4}{2^{2n+2k}}\; (\hat{g_M}(s))^2\nonumber\\
 & = &  2^{n+2} \frac{1}{N} \sum_{M\in \CM}
\mathop{\sum_{s: h(s)=k}}_{k\; odd}
\frac{1}{2^{2k}} (\hat{g_M}(s))^2
\end{eqnarray}
Putting (\ref{1}) and (\ref{intermed}) together we have
\begin{equation} \label{ineq}
(1-4\epsilon)^2 \leq 2^{2n} \frac{1}{N} \sum_{M\in \CM}
\mathop{\sum_{s: h(s)=k}}_{k\; odd}
\frac{1}{2^{2k}} (\hat{g_M}(s))^2
\end{equation}
We now relate the Fourier coefficients of $g_M$ with those of $g$. Note first that if $M$ is an $(n\times 2n)$ matrix, $x\in\zo^{2n}$ and  $s\in\zo^n$ then
\[ (Mx)^T\cdot  s = (x^T M^T)\cdot s = x^T \cdot(M^T s) \]
By the definition of $g_M$, its Fourier coefficients are
\[
\hat{g}_M(s) = \frac{1}{2^n}\sum_y g_M(y)(-1)^{y^T\cdot s} =
\frac{1}{2^n|A|} \Big(|\{x\in A | (Mx)^T\cdot s =0\}|- |\{x\in A |
(Mx)^T \cdot s=1\}|\Big).  \]
Let $s_M = M^T s$ and note that $h(s_M) = 2h(s)$. Then,
\begin{eqnarray*}
\hat{g}(s_M) & = & \frac{1}{2^{2n}} \sum_{x\in \zo^{2n}}
g(x)(-1)^{x^T\cdot s_M} =
\frac{1}{2^{2n}|A|} \Big(|\{x\in A |\; x^T \cdot s_M =0\}|- |\{x\in A |\;
x^T \cdot s_M =1\}|\Big) \\
& = & \frac{1}{2^{2n}|A|} \Big(|\{x\in A |\; (Mx)^T\cdot s =0\}|- |\{x\in A |\;
(Mx)^T\cdot s  =1\}|\Big) \\
& = & \frac{1}{2^n}\hat{g}_M(s)
\end{eqnarray*}
Inequality (\ref{ineq}) now becomes
\begin{eqnarray}\label{4}
(1-4\epsilon)^2 & \leq & 2^{2n} \frac{1}{N} \sum_{M\in \CM}
\mathop{\sum_{s: h(s)=k}}_{k\; odd}
\frac{1}{2^{2k}} (\hat{g_M}(s))^2 \nonumber\\
& = &
2^{2n}\frac{1}{N}\sum_{M\in \CM} \mathop{\sum_{s: h(s)=k}}_{k\; odd}
\frac{1}{2^{2k}}\; 2^{2n}\; (\hat{g}(s_M))^2  \nonumber\\
& = &  2^{4n} \frac{1}{N} \sum_{M\in \CM}
\mathop{\sum_{s: h(s)=k}}_{k\; odd}
\frac{1}{2^{2k}} (\hat{g}(s_M))^2
\end{eqnarray}
In the above expression we first sum over all matchings and then over the string $s$. In what follows we will try to change the order of the summation. Note that in the above expression when $h(s)$ is odd, $h(s_M)= 2 \mod 4$.
For any $k =2 \mod{4}$ we define $\gamma_k$ as follows:\\
Let $z\in\zo^{2n}$ be any string of hamming weight $k$ and $M$ be a random matching. Then
\[ \gamma_k = \prob_M[\exists s\; \mbox{ s.t. } z=s_M]
\]
Note that this probability depends only on $k$ and not on the specific string $z$. For any even number $t\geq 2$, let $N(t)$ be the number of perfect matchings on $[t]$. Then,
$$ N(2)=1,\;\; N(t) = (t-1)N(t-2).$$
It is not hard to see that the expression for $\gamma_k$ is
\[ \gamma_k = \frac{N(k) N(2n-k)}{N(2n)} = \frac{(k-1)(k-3)\cdot\ldots\cdot 1}{(2n-1)(2n-3)\cdot\ldots\cdot (2n-k+1)} \leq \left( \frac{k}{2n}\right)^{k/2}\]
We now rewrite inequality (\ref{4}) after changing the order of the summation
\begin{eqnarray*}
(1-4\epsilon)^2 & \leq &
2^{4n} \mathop{\sum_{z: h(z)=k}}_{k = 2 (mod 4)} \frac{1}{2^k}\;\gamma_k\; (\hat{g}(z))^2
\end{eqnarray*}
and hence
\begin{eqnarray*}
1 & \leq & \frac{1}{(1-4\epsilon)^2} \;2^{4n} \mathop{\sum_{z: h(z)=k}}_{k = 2 (mod 4)} \frac{1}{2^k}\;\gamma_k\; (\hat{g}(z))^2 \\
& = & \sum_{k = 2 (mod 4)} \frac{1}{(1-4\epsilon)^2} \;2^{4n} \sum_{z: h(z)=k} \frac{1}{2^k}\;\gamma_k\; (\hat{g}(z))^2
\end{eqnarray*}
From the above inequality, it is easy to see that there exists a $k$ such that
\begin{eqnarray*}
\frac{1}{(1-4\epsilon)^2}\; 2^{4n} \sum_{z: h(z)=k} \frac{1}{2^k} \gamma_k(\hat{g}(z))^2 & \geq & \frac{1}{2^{k/2}}\Big/\sum_{r = 2 (mod 4)} \frac{1}{2^{r/2}}
\end{eqnarray*}
otherwise sum over $k = 2\mod 4$ to get a contradiction. Moreover,
$$\sum_{r = 2 (mod 4)} \frac{1}{2^{r/2}} \leq \frac{2}{3}$$
and hence for that $k$
\begin{eqnarray*}
2^{4n} \sum_{z: h(z)=k} \frac{1}{2^{k/2}} \gamma_k(\hat{g}(z))^2 & \geq & \frac{3(1-4\epsilon)^2}{2}.
\end{eqnarray*}
We choose small enough $\epsilon$ such that $\frac{3(1-4\epsilon)^2}{2}\geq 1$ and then we rewrite the above inequality as
\begin{eqnarray} \label{delta}
\frac{2^{4n}}{|A|^2} \sum_{z: h(z)=k} \gamma_k|A|^2(\hat{g}(z))^2 & \geq & 2^{k/2}
\end{eqnarray}
Let $\delta = (\gamma_k)^{1/k}$. Note that $0 \leq \delta \leq 1$. By the KKL inequality (lemma 3) \cite{KKL88}, we know that
\[ \sum_{z: h(z)=k} \delta^k |A|^2(\hat{g}(z))^2  \leq \left(\frac{|A|}{2^{2n}}\right)^{\frac{2}{1+\delta}}.
\]
Hence, inequality (\ref{delta}) becomes
\begin{eqnarray*}
2^{k/2}  & \leq & \frac{2^{4n}}{|A|^2} \left(\frac{|A|}{2^{2n}}\right)^{\frac{2}{1+\delta}} =
\left(\frac{|A|}{2^{2n}}\right)^{\frac{-2\delta}{1+\delta}}
\leq \left(\frac{|A|}{2^{2n}}\right)^{-2\delta}
\end{eqnarray*}
and therefore
\begin{eqnarray}\label{gamma}
\frac{|A|}{2^{2n}} \leq \left( 2^{k/2} \right)^{-1/2\delta}
 = 2^{-k/4\delta}
\end{eqnarray}
We know that $\gamma_k \leq \left(\frac{k}{2n} \right)^{k/2}$ and $k\geq 2$, so
\[ \frac{k}{4\delta} = \frac{k}{4(\gamma_k)^{1/k}} \geq \frac{k}{4\left(\frac{k}{2n}\right)^{1/2}} = \frac{\sqrt{2nk}}{4} \geq \frac{\sqrt{n}}{2}.
\]
From inequality (\ref{gamma}) we conclude that
\begin{eqnarray*}
\frac{|A|}{2^{2n}} & \leq & 2^\frac{-\sqrt{n}}{2} \\
|A|  & \leq & 2^{2n-\Omega(\sqrt{n})}
\end{eqnarray*}
Since the size of any $|A|$ cannot be more than $2^{2n-\Omega(\sqrt{n})}$, it means that there are at least $2^{\Omega(\sqrt{n})}$ different sets $A$. In other words, there are at least $2^{\Omega(\sqrt{n})}$ different messages that Alice sends and therefore the length of her message is at least $\Omega(\sqrt{n})$.
\end{proof}

\newcommand{\etalchar}[1]{$^{#1}$}
\bibliographystyle{plain}

\end{document}